\documentclass[12pt]{article}
\usepackage[a4paper,top=4cm,bottom=4cm,left=3cm,right=3cm]{geometry}
\usepackage[colorlinks=true,citecolor=blue]{hyperref}
\usepackage{mathptmx, amsmath, amssymb, amsfonts, amsthm, mathptmx, enumerate, color,mathrsfs}
\usepackage[dvipsnames]{xcolor}
\usepackage[mathscr]{eucal}
\usepackage{graphicx}
\graphicspath{{plots/}}
\usepackage{caption} 
\usepackage{subcaption}
\usepackage[percent]{overpic}
\usepackage{pict2e}

\usepackage{multirow}
\usepackage{booktabs}
\usepackage{epstopdf}
\usepackage{multicol}
\usepackage{algorithm}
\usepackage{algorithmic}
\usepackage{epstopdf}
\usepackage{amsmath}
\usepackage{bm}
\usepackage[most]{tcolorbox}
\usepackage{tikz}
\usepackage{etoolbox} 
\usepackage{listofitems} 

\tikzset{>=latex} 
\colorlet{myred}{red!80!black}
\colorlet{myblue}{blue!80!black}
\colorlet{mygreen}{green!60!black}
\colorlet{mydarkred}{myred!40!black}
\colorlet{mydarkblue}{myblue!40!black}
\colorlet{mydarkgreen}{mygreen!40!black}
\tikzstyle{node}=[very thick,circle,draw=myblue,minimum size=22,inner sep=0.5,outer sep=0.6]
\tikzstyle{connect}=[->,thick,mydarkblue,shorten >=1]
\tikzset{ 
  node 1/.style={node,mydarkgreen,draw=mygreen,fill=mygreen!25},
  node 2/.style={node,mydarkblue,draw=myblue,fill=myblue!20},
  node 3/.style={node,mydarkred,draw=myred,fill=myred!20},
}
\def\nstyle{int(\lay<\Nnodlen?min(2,\lay):3)} 

\newcommand{\norm}[1]{\left\lVert#1\right\rVert}
\def\vconst{v_\mathrm{const}}
\def\mmqs{\mathrm{mm^2/s}}
\def\mms{\mathrm{mm/s}}
\def\mutest{\bm\mu^\mathrm{test}}
\def\mutrain{\bm\mu^\mathrm{train}}

\theoremstyle{definition}

\numberwithin{equation}{section}

\title{\large\sc A data-driven Reduced Order Method for parametric optimal blood flow control: application to coronary bypass graft}
\author{
\normalsize{Caterina Balzotti} \thanks{SISSA, International School for Advanced Studies, Mathematics Area, mathLab, via Bonomea 265, I-34136 Trieste, Italy (\href{mailto:cbalzott@sissa.it}{cbalzott@sissa.it}, \href{mailto:psiena@sissa.it}{psiena@sissa.it}, \href{mailto:mgirfogl@sissa.it}{mgirfogl@sissa.it}, \href{mailto:grozza@sissa.it}{grozza@sissa.it})}
\and{\normalsize{Pierfrancesco Siena}}\footnotemark[1]
\and{\normalsize{Michele Girfoglio}}\footnotemark[1]
\and{\normalsize{Annalisa Quaini}}\thanks{University of Houston, Department of Mathematics, 3551 Cullen Blvd, 77204, Houston TX, USA (\href{mailto:quaini@math.uh.edu}{quaini@math.uh.edu})}
\and{\normalsize{Gianluigi Rozza}}\footnotemark[1]
}
\date{\vspace{-1cm}}

\begin{document}

\maketitle

\begin{abstract}
We consider an optimal flow control problem in a patient-specific coronary artery bypass graft with the aim of matching
the blood flow velocity with given measurements as the Reynolds number varies in a physiological range. 
Blood flow is modelled with the steady incompressible Navier-Stokes equations.
The geometry consists in a stenosed left anterior descending artery where a single bypass is performed with the right internal thoracic artery.  
The control variable is the unknown value of the normal stress at the outlet boundary, which is need for a correct set-up of the outlet boundary condition. For the numerical solution of the parametric optimal flow control problem, we develop a data-driven reduced order method that combines proper orthogonal decomposition (POD) with neural networks. We present numerical results showing that our data-driven approach 
leads to a substantial speed-up with respect to a more classical POD-Galerkin strategy proposed in \cite{zakia2021}, while having comparable accuracy. 

\end{abstract}
 

\vskip 1cm
\centerline{Dedicated to the memory of Roland Glowinski}

\section{Introduction}
It is well-known that cardiovascular disease is one of the leading causes of death in the world. This has motivated a large body of literature on mathematical and numerical modeling for blood flow problems
since computational simulations in patient-specific geometries can assist medical doctors in clinical and surgical practice. 
This paper focuses on blood flow in coronary arteries when poor blood perfusion requires a surgical treatment known as 
coronary artery bypass graft (CABG). For recent computational work on this topic, we refer to, e.g., \cite{ballarin2016,ballarin2017,sienadata,sienafast,shapedesign2006,mathapproach2006,rozzarealtimered2005}.  

Computational hemodynamics in realistic geometries is challenging for several reasons. Here, we tackle two main challenges: (i) the high computational cost and (ii) the enforcement of boundary conditions that would lead to realistic flow. To deal with point (i), we develop a computationally efficient data-driven reduced order method (ROM). Our goal is not only to reduce drastically the computational time required by full order simulations, but also to 
achieve a speed-up with respect to a 
traditional reduced method (POD-Galerkin) used in \cite{zakia2021}. As for point (ii), we set up an optimal control framework where the control variable is the unknown normal stress at the outlet boundary. Typically, one directly  imposes boundary conditions with specified flow waveform at the inlets and constant values of pressure or velocity gradient
at the outlets.
See, e.g., \cite{sienadata,sienafast}. However, while this approach allows to take into account patient specific flow profiles, the non-physical outlet boundary conditions lead to inaccurate numerical approximations. An alternative is to employ surrogate models to enforce boundary conditions \cite{girfogliononintrusive,girfoglpodirom,fevola2021}. This results in increased accuracy, but requires a very large number of parameters which increases the model complexity too.

Our optimal control flow problem involves the minimization of an objective functional measuring the mismatch between computed and measured velocity fields and solves for the unknown control variable (i.e., the normal stress at the outlet boundary) with the constraint of the governing equations (i.e., the parameterized Navier-Stokes equations) \cite{gunzburger2002,quarteroni2014,hinze2009,quarteronioptimal,hounumerical1999,dedeoptimal2007}. 
The formulation of this optimal control problem was introduced in \cite{zakia2021}, with the assumption that the velocity data comes from 
magnetic resonance imaging four-dimensional flow. In particular, we aim to minimize in a least-square sense the difference between numerical and measured velocity \cite{tiago2017,Romarowski2018,Koltukluoglu} in a patient specific geometry reconstructed from a computed tomography scan. 
In order to consider various hemodynamics scenarios, we let the Reynolds number vary in a given range. 
A finite element approach for the problem under consideration features high computational cost. To contain such cost, we adopt a data-driven ROM that combines proper orthogonal decomposition with neural networks (NNs) known as POD-NN \cite{Hesthaven2018}. 
The POD algorithm extracts the dominant modes that form a basis for the reduced space and NNs are used to interpolate the modal coefficients for each variable. During the so-called offline stage of the ROM, the POD algorithm is executed and the feedforward neural networks are trained. This is an expensive stage due to the high dimension of the full order model (i.e., the finite element method), which however needs to be performed only once. The so-called online phase requires only the evaluation of the modal coefficients through the trained NNs for each new Reynolds number of interest. The complete decoupling between offline/online phases together with a data-driven approach leads to a considerable computational efficiency, thereby allowing to solve the optimization problem for several hemodynamics scenarios at a greatly reduced cost.

The remainder of this paper is structured as follows. In Section \ref{ocp}, we report the mathematical details about the formulation of the continuous optimal control problem, together with the patient-specific geometry of the CABG. Section \ref{numericsection} describes the finite element discretization of our problem and introduces the data-driven reduced order framework. Numerical results are shown in Section \ref{sec:num_res}. Conclusions and future perspectives are drawn in Section \ref{sec:concl}.

\section{Optimal control problem}
\label{ocp}
In this section, we introduce the mathematical details for optimal control flow problems in the context of a patient-specific CABG geometry. 

In general, an optimal control flow problem is composed of three elements:
\begin{itemize}
    \item[-] an \emph{objective functional} to optimize;
    \item[-] a \emph{control variable} to choose in order to minimize the objective functional;
    \item[-] the \emph{fluid flow model}, which represents a set of constraints for the optimization step.
\end{itemize}
More details can be found in, e.g., \cite{gunzburger2002,bewley2001,Gad-elHak2003}. In our specific constrained optimization problem, the aim is to minimize an objective functional representing the distance between computational results and clinical data.

This section is organized as follows. 
First, we provide some information about the spatial domain. Then, we briefly describe the governing equations for blood flow in large vessels, i.e., the incompressible Navier-Stokes equations. Finally, 
we cast the mathematical problem into a saddle-point formulation, which we will argue is easier to handle. 

\subsection{The geometry}
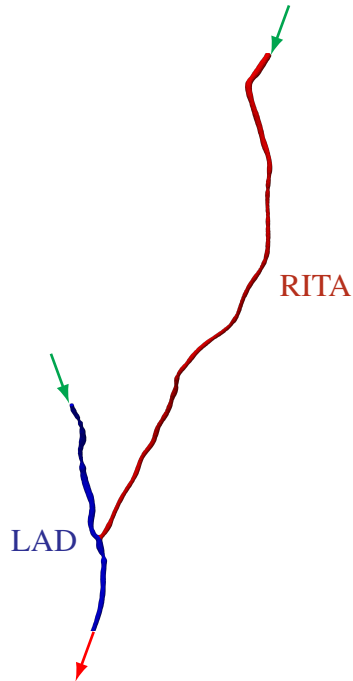
\begin{figure}[h]
    \centering
    \vspace{0.5cm}
    \begin{overpic}[width=0.18\columnwidth, trim = 0cm -9cm 0cm 0cm]{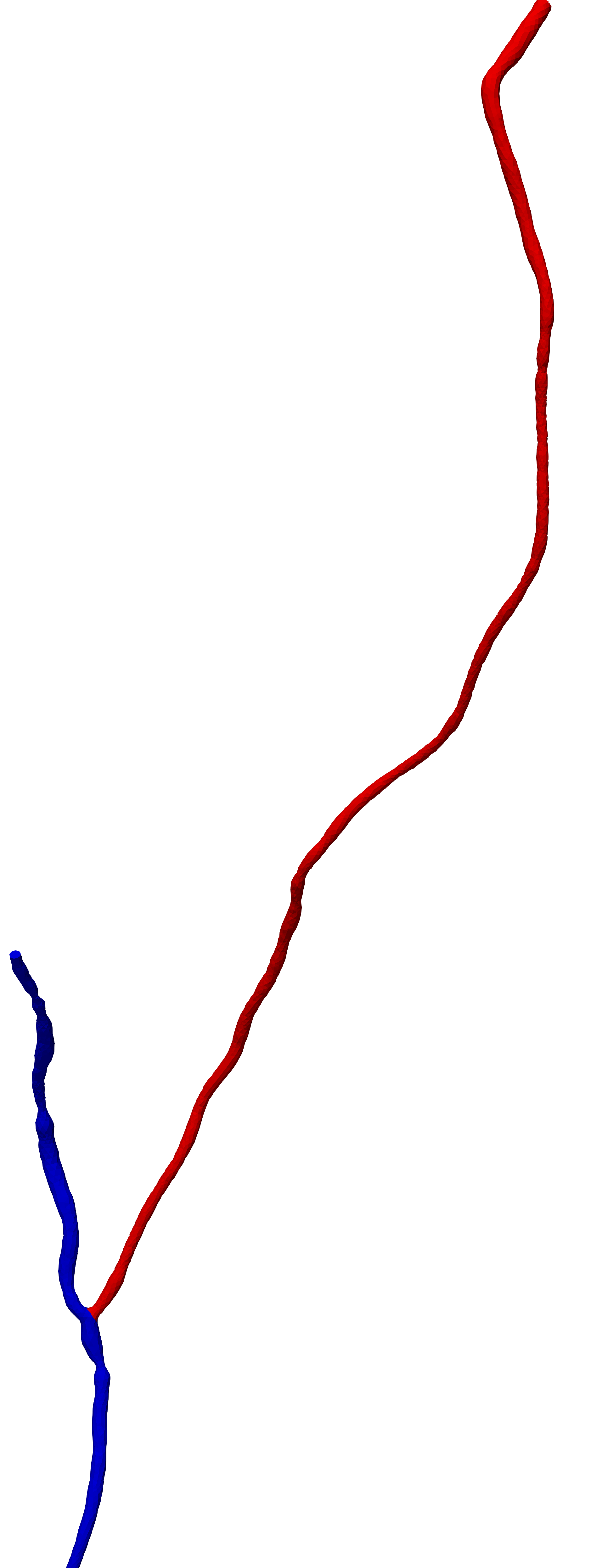}
    \put(-9,18){\textcolor{Blue}{LAD}}
    \put(35,60){\textcolor{BrickRed}{RITA}}
    \linethickness{1pt}\put(-2.5,50.5){\color{Green}\vector(0.08,-0.22){3}}
    \linethickness{1pt}\put(36.5,108){\color{Green}\vector(-0.08,-0.22){3}}
    \linethickness{1pt}\put(4.5,4.6){\color{red}\vector(-0.08,-0.22){3}}
    \end{overpic}
    \caption{The spatial domain: a coronary artery bypass graft (CABG) performed with the right internal thoracic artery (RITA) on the left anterior descending artery (LAD). The arrows indicate the flow direction.
    }
    \label{fig:geometry}
\end{figure}

The CABG geometry used for this work is shown in Figure \ref{fig:geometry}. A post-surgery computed tomography (CT) scan was provided by the \textit{Sunnybrook Health Science Centre} in \textit{Toronto}. The patient exhibited at least three stenoses in the coronary system ranging from mild to severe. We will consider only one of such stenoses to test our methodology. The bypass of interest is performed with the right internal thoracic artery (RITA) on the left anterior descending artery (LAD). See Figure \ref{fig:geometry}.
The approach used to extract the domain is based on \textit{Visualization Toolkit} (VTK) and \textit{Vascular Modeling Toolkit} (VMTK) Python libraries \cite{schroeder2006,antiga2003} and it is accurately described in \cite{ballarin2016,ballarin2017}.

\subsection{Problem formulation}
Let us consider the computational domain $\Omega \subset \mathbb{R}^3$ in Figure \ref{fig:geometry} and denote by $\partial \Omega$ its boundary. More precisely, the boundary consists of:
 \begin{itemize}
     \item two inlets denoted by $\Gamma_{\text{inlet}}$, where the flux enters the RITA and the LAD;
     \item one outlet $\Gamma_{\text{outlet}}$, where the blood leaves the LAD;
     \item the rigid vessel wall $\Gamma_{\text{wall}}$.
 \end{itemize}
Due to the radii of the coronary arteries, the apparent viscosity of the blood can be considered constant.  Therefore, we adopt a Newtonian model (see also \cite{torii2003,abbasian2018}). 

Let $\bm \mu \in \mathcal{P}$ be a set of physical parameters belonging to a given parameter space $\mathcal{P}$. 
The parameterized, steady-state incompressible Navier-Stokes equations describing the blood flow are given by:
\begin{equation}
	\begin{cases} 
     - \nu \Delta \bm{v}(\bm \mu) + (\bm{v}(\bm \mu) \cdot \nabla) \bm{v}(\bm \mu) + \nabla p(\bm \mu)  = 0,  & \quad \text{in} \quad \Omega, \\
	\nabla \cdot \bm{v} = 0, & \quad \text{in} \quad \Omega,
	\end{cases}  \label{N-S}
\end{equation}
where $\bm v(\bm \mu)$ is the velocity field, $p(\bm\mu)$ the blood pressure, and $\nu$ the constant viscosity. Problem \eqref{N-S} is endowed with a non-homogeneous Dirichlet condition at the inlet, a non-homogeneous Neumann condition at the outlet, and a 
no-slip condition on the wall:
\begin{equation}
    \begin{cases}
     \bm v (\bm \mu) = \bm v_{\text{inlet}} (\bm \mu),  & \quad \text{on} \quad \Gamma_{\text{inlet}}, \\
     -\nu(\nabla\bm v (\bm \mu))\bm n +p(\bm\mu)\bm n = \bm u (\bm \mu),  & \quad \text{on} \quad \Gamma_{\text{outlet}}, \\
     \bm v (\bm \mu) = \boldsymbol{0},  & \quad \text{on} \quad \Gamma_{\text{wall}},
    \end{cases} \label{N-S-bc}
\end{equation}
where $\bm v_{\text{inlet}}(\bm \mu)$ is a given inlet velocity profile, $\bm n$ is the outward unit normal vector, and $\bm u(\bm \mu)$ is the unknown control variable.

Let us introduce the Hilbert spaces $V = V(\Omega)$ and $Q = Q(\Omega)$, with the corresponding dual spaces $V^*$ and $Q^*$. 
Let $S=V\times Q$ be the self-adjoint space (i.e., $S=S^*$) of the state variables $(\bm v (\bm \mu), p (\bm \mu))$. Moreover, let 
$U = U(\Gamma_{\text{outlet}})$ be the Hilbert space the control $\bm u$ belongs to. In order to state the weak formulation of problem \eqref{N-S}-\eqref{N-S-bc}, we need to introduce forms $a:V\times V^* \rightarrow \mathbb{R}$, $e:V\times V \times V^* \rightarrow \mathbb{R}$, $b:Q^*\times V \rightarrow \mathbb{R}$ and $c:U\times V^* \rightarrow \mathbb{R}$:
\begin{align*}
    a(\bm v, \bm w; \bm \mu)  &= \nu \int_{\Omega} \nabla \bm v(\bm \mu) \cdot\nabla \bm w d\Omega, \quad e(\bm v,\bm v,\bm w;\bm \mu) = \int_{\Omega} (\bm v(\bm \mu) \cdot \nabla)\bm v(\bm \mu) \cdot \bm w d\Omega,\\ 
b(q, \bm v; \bm \mu) &= -\int_{\Omega}q(\nabla \cdot \bm v(\bm \mu))d\Omega, \quad c(\bm u, \bm w; \bm \mu) = -\int_{\Gamma_{outlet}}\bm u(\bm \mu) \cdot \bm w d\Gamma.
\end{align*}
Then, the weak formulation can be written as:
Given $\bm \mu \in \mathcal{P}$, find the solution $\bm s (\bm \mu) = (\bm v (\bm \mu),p(\bm \mu))$ $\in S$ 
such that:
\begin{equation}
    \begin{cases}
    a(\bm v, \bm w; \bm \mu) +e(\bm v,\bm v,\bm w;\bm \mu) + b(p, \bm w; \bm \mu) + c(\bm u, \bm w; \bm \mu) = 0, & \quad \forall \bm w \in V^*, \\
    b(q, \bm v; \bm \mu) = 0, & \quad \forall \bm q \in Q^*.
    \end{cases} \label{N-S-debole}
\end{equation}

The objective functional chosen for this work measures the distance between $\bm v$ and the clinically measured 
velocity $\bm v_{\text{m}}$ as:
\begin{equation}
    \begin{split}
        \mathcal{I} (\bm v, \bm u; \bm \mu) &= \frac{1}{2}\int_{\Omega}|\bm v (\bm \mu)-\bm v_{\text{m}}|^2 d\Omega + \frac{\alpha}{2}\int_{\Gamma_{outlet}}|\bm u (\bm \mu)|^2 d \Gamma.
    \end{split}
    \label{func}
\end{equation}
where $\alpha>0$ is a constant penalization parameter. 
Let us introduce Hilbert space $G \supseteq S$, bilinear forms $m:G\times G\rightarrow \mathbb{R}$ and $n:U\times U \rightarrow \mathbb{R}$ such that the cost functional \eqref{func} can be written as:
\begin{equation}
    \mathcal{I} (\bm v, \bm u; \bm \mu) = \frac{1}{2} m( \bm v (\bm \mu)-\bm v_{\text{m}}, \bm v (\bm \mu)-\bm v_{\text{m}})+\frac{\alpha}{2}n(\bm u (\bm \mu),\bm u (\bm \mu)).
    \label{func1}
\end{equation}

The optimal control flow problem reads:
\begin{tcolorbox}[colback=lightgray!5!white,colframe=lightgray!75!black]
Given $\bm \mu \in \mathcal{P}$, find $(\bm v(\bm \mu), p(\bm \mu),\bm u (\bm \mu))\in V \times Q \times U$ such that the objective functional \eqref{func1} is minimized under the constrain \eqref{N-S-debole}.
\end{tcolorbox}

\subsection{Saddle-point structure}
A classical approach to treat non-linear optimal control problems is the adjoint-based Lagrangian method. 
Here, we briefly describe the procedure adopted to cast the constrained optimization problem with the well-known saddle-point theory. The reader interested in more details is referred to, e.g., \cite{bochev2009,boffi2013,quarteroni2014}. 

Let us introduce $X = V \times Q \times U$ and denote by $\bm x = (\bm v,p,\bm u)\in X$ 
the state and the control variables of the optimal control problem. In addition, let $\bm y = (\bm y_{\bm v},\bm y_p, \bm y_{\bm u})\in X$. We define $\mathcal{A}:X\times X \rightarrow \mathbb{R}$ as: 
 \begin{equation*}
     {\mathcal{A}(\bm x, \bm y;\bm \mu) = m(\bm v(\bm \mu),\bm y_{\bm v}) + \alpha n(\bm u(\bm \mu),\bm y_{\bm u}), \quad \forall \bm x, \bm y \in X.}
 \end{equation*}
Moreover, let $\mathcal{B}:X\times S^* \rightarrow \mathbb{R}$ be the operator related to the linear part of \eqref{N-S-debole}, such that we can rewrite \eqref{N-S-debole} as:
\begin{equation}
    \mathcal{B}(\bm x, \bm z ;\bm \mu)+ e(\bm v,\bm v, \bm w; \bm \mu) = 0, \quad \forall \bm z \in S^*,
    \label{eq_with_B}
\end{equation}
with $\bm z = (\bm w, q)$. 
Finally, let $\bm h(\bm\mu)\in X^*$ be such that
\begin{equation*}
     {\left \langle \bm h(\bm\mu),\bm y \right \rangle = m(\bm v, \bm v_{\text{m}} ). \quad \forall \bm y \in X,}
 \end{equation*}
Therefore, a new functional can be introduced:
\begin{equation}
\begin{split}
    {J(\bm x;\bm \mu)} &= {\frac{1}{2}\mathcal{A}(\bm x, \bm x;\bm \mu)- \left \langle \bm h (\bm\mu),\bm x \right \rangle} = \mathcal{J}(\bm v, \bm u;\bm \mu) - \frac{1}{2}m(\bm v_{\text{m}},\bm v_{\text{m}}).
\end{split} \label{func_new}
\end{equation}
Since $m(\bm v_{\text{m}},\bm v_{\text{m}})$ is constant, an equivalent form of the optimal control flow problem is:

\begin{tcolorbox}[colback=lightgray!5!white,colframe=lightgray!75!black]
Given $\bm \mu \in \mathcal{P}$, find $\bm x (\bm \mu) =(\bm v(\bm \mu), p(\bm \mu),\bm u (\mu)) \in X$ such that the objective functional  \eqref{func_new} is minimized under the constrain \eqref{eq_with_B}.
\end{tcolorbox}

In order to obtain an unconstrained formulation, let us define the Lagrangian $\mathcal{L}:X\times S^*\rightarrow \mathbb{R}$:
\begin{equation*}
    \mathcal{L}(\bm x, \bm z;\bm \mu) = J(\bm x;\bm \mu) +  \mathcal{B}(\bm x, \bm z ;\bm \mu)+ e(\bm v,\bm v, \bm w; \bm \mu).
\end{equation*}
The saddle-point structure is obtained with the optimality condition: $\nabla \mathcal{L}(\bm x, \bm z;\bm \mu) [\bm y, \bm k]= 0, \forall \bm y \in X, \forall \bm k = (\bm k_{\bm w}, \bm k_{q}) \in S^*$.
This leads to the following formulation:
\begin{tcolorbox}[colback=lightgray!5!white,colframe=lightgray!75!black]
Given $\bm \mu \in \mathcal{P}$, find $(\bm x(\bm \mu), \bm z(\bm \mu))\in X \times S^* $ such that:
\begin{equation}
    \begin{cases}
      \mathcal{A}(\bm x, \bm y;\bm \mu) + \mathcal{B}(\bm y, \bm z ;\bm \mu)+ e(\bm y_{\bm v},\bm v, \bm w; \bm \mu)+ e(\bm v,\bm y_{\bm v}, \bm w; \bm \mu) =  \left \langle \bm{h}(\bm\mu),\bm y \right \rangle, \quad \forall \bm y \in X, \\
      \mathcal{B}(\bm x, \bm k ;\bm \mu)+ e(\bm v,\bm v, \bm k_{\bm w}; \bm \mu) = 0, \quad \forall \bm k \in S^*.
    \end{cases} \label{optimal-control-saddle-point}
\end{equation}
\end{tcolorbox}
Existence and uniqueness of the solution to the above problem is ensured by Brezzi's theorem (see \cite{boffi2013} for details).

\section{Numerical discretization}
\label{numericsection}
In this section, we introduce the numerical methods to find the fluid state variables, the control variable and the adjoint variables.
To find the full order solution of problem \eqref{optimal-control-saddle-point},
we adopt an \textit{optimize-then-discretize} approach. This procedure formalizes the continuous problem and then discretizes the optimal control system. We choose
a finite element method for the discretization and use a \textit{one-shot} method to solve directly the resulting equations. See, e.g., \cite{schulz2009,gunzburger2002} for details.

For the reduction step, we opt for 
a \textit{data-driven}, \textit{non-intrusive} ROM. More precisely, 
we employ the POD-NN method \cite{Hesthaven2018}. Our aim is to compare our results with those obtained with an intrusive POD-Galekin procedure in \cite{zakia2021}.

\subsection{The full order model}

We consider a mesh for domain $\Omega$ with size $h$.
Let us consider spaces $S_h=V_h \times Q_h \subset S$ and $X_h = S_h \times U_h \subset X$. 
The full order model (FOM) is:
\begin{tcolorbox}[colback=lightgray!5!white,colframe=lightgray!75!black]
Given $\bm \mu \in \mathcal{P}$, find $(\bm x_h(\bm \mu), \bm z_h(\bm \mu))\in X_h \times S_h^* $ such that:
\begin{equation}
\small
    \begin{cases}
      \mathcal{A}(\bm x_h, \bm y_h;\bm \mu) + \mathcal{B}(\bm y_h, \bm z_h ;\bm \mu)+ e(\bm y_{\bm v_h},\bm v_h, \bm w_h; \bm \mu)+ e(\bm v_h,\bm y_{\bm v_h}, \bm w_h; \bm \mu) =  \left \langle \bm{h}(\bm\mu),\bm y_h \right \rangle, \quad \forall \bm y_h \in X_h, \\
      \mathcal{B}(\bm x_h, \bm k_h ;\bm \mu)+ e(\bm v_h,\bm v_h, \bm k_{\bm w_h}; \bm \mu) = 0, \quad \forall \bm k_h \in S_h^*.
    \end{cases} \label{optimal-control-saddle-point-discr}
\end{equation}
\end{tcolorbox}

We refer to eq.~\eqref{optimal-control-saddle-point-discr} as the \textit{truth problem}, which has dimension $\displaystyle{{N}=\sum_{\delta}{N}_{\delta}}$, with ${N}_{\delta}$ the dimension of the solution space for variable $\delta=\bm v, p, \bm u, \bm w, q$. 

In order to introduce the matrix formulation of the 
truth problem, let $\{ \phi_i\}_{i=1}^{{N}_{\bm v}}$, $\{ \psi_k\}_{k=1}^{{N}_{p}}$ and $\{ \sigma_l\}_{l=1}^{{N}_{\bm u}}$ be bases for $V_h$, $Q_h$ and $U_h$, respectively. The state and control variables can be written as linear combination of the basis functions:
\begin{equation*}
    \bm v_h(\bm \mu) = \sum_{i=1}^{{N}_{\bm v}} 
    \bar{v}_i(\bm \mu) \phi_i, \quad p_h(\bm \mu) = \sum_{k=1}^{{N}_{p}} 
    \bar{p}_k(\bm \mu) \psi_k, \quad  \bm u_h(\bm \mu) = \sum_{l=1}^{{N}_{\bm u}} 
    \bar{u}_l(\bm \mu)\sigma_l,
\end{equation*}
where $\bm{\bar v}(\mu) = [ \bar{v}_1, \dots, \bar{v}_{{N}_{\bm v}}]^T$, $\bm{\bar p}(\mu) = [ \bar{p}_1, \dots, \bar{p}_{{N}_p}]^T$ and $\bm{\bar u}(\mu) = [ \bar{u}_1, \dots, \bar{u}_{{N}_{\bm u}}]^T$ are the vectors of the coefficients.
Taking the basis functions as test functions for problem \eqref{optimal-control-saddle-point-discr},
we introduce the matrices associated with the forms in \eqref{N-S-debole}:
$$
(A(\bm \mu))_{ij}=a(\phi_i,\phi_j;\bm \mu), \quad (B(\bm \mu))_{ik}= b(\psi_k,\phi_i;\bm \mu), \quad (C(\bm \mu))_{il} =  c(\sigma_l,\phi_i;\bm \mu), $$
$$
(M(\bm \mu))_{ij} =  m(\phi_i,\phi_j;\bm \mu), \quad N(\bm \mu)_{lr} = n(\sigma_r,\sigma_l;\bm \mu),
$$
with $1 \le i,j \le {N}_{\bm v}$, $1 \le k \le {N}_p$ and $1 \le l,r \le {N}_{\bm u}$.
Furthermore, we define:
$$
(E(\bm{\bar v}(\bm \mu);\bm \mu))_{ij} = \sum_{k=1}^{{N}_v} 
\bar{v}_k(\bm \mu) e(\phi_k,\phi_j,\phi_i;\bm \mu), \quad (\tilde{E}(\bm{\bar w}(\bm \mu);\bm \mu))_{ij} = \sum_{k=1}^{{N}_v}
\bar{w}_k(\bm \mu) e(\phi_k,\phi_j,\phi_i;\bm \mu),
$$
where $\bm{\bar w}(\mu) = [ \bar{w}_1, \dots, \bar{w}_{{N}_v}]^T$ is the vector of the coefficients of the adjoint velocity and $\bm{\bar q}(\mu) = [ \bar{q}_1, \dots, \bar{q}_{{N}_p}]^T$ is the vector of the coefficients for the adjoint pressure.

The algebraic formulation of the truth problem can be written as:
\begin{equation}
\small
    \begin{Bmatrix} 
    M(\bm \mu) + \tilde{E}(\bm{\bar w}(\bm \mu);\bm \mu) & 0 & 0 & A(\bm \mu)+ E(\bm{\bar v}(\bm \mu);\bm \mu) & B(\bm \mu) \\
    0 & 0 & 0 & B^T(\bm \mu) & 0\\
    0 & 0 & \alpha N(\bm \mu) & C(\bm \mu) & 0 \\
    A(\bm \mu) + E(\bm{\bar v}(\bm \mu);\bm \mu) & B^T(\bm \mu) & C(\bm \mu) & 0 & 0 \\
    B(\bm \mu) & 0 & 0 & 0 & 0 
    \end{Bmatrix}
    \begin{Bmatrix}
    \bm{\bar v} (\bm \mu) \\
    \bm{\bar p}(\bm \mu) \\
    \bm{\bar u}(\bm \mu) \\
    \bm{\bar w}(\bm \mu) \\
    \bm{\bar q}(\bm \mu) \\
    \end{Bmatrix} = 
    \begin{Bmatrix}
    \bm h (\bm \mu) \\
    \bm 0 \\
    \bm 0 \\
    \bm 0 \\
    \bm 0 \\
    \end{Bmatrix}.
    \label{matrixform}
\end{equation}

One way to ensure uniqueness of the solution for the above system is to employ inf-sup stable finite element pairs for  velocity and pressure.
See, e.g., \cite{ballarinsupremizer2015,ali2020} for more details.


\subsection{The reduced order model}\label{sec:rom}
Like many other ROM techniques, 
the POD-NN method relies on the \emph{offline}-\emph{online} paradigm, i.e., computationally intensive simulations are carried out on powerful computational facilities during the offline phase and computationally cheap calculations for every new parameter of interest are run on any device during the online phase. More specifically:
\begin{itemize}
	\item[-] during the \emph{offline} phase: a set of high-fidelity solutions is collected for a wide range of parameter values and a reduced basis for the space of the reduced solutions is extracted via POD. In this stage, the training of the neural networks is performed in order to establish the relationship between parameters and coefficients of the reduced solutions (modal coefficients).
    Due to the large number of degree of freedom, the offline phase is computationally expensive. However, it is carried out only once. 
	\item[-] during the \emph{online} phase: the modal coefficients for every new parameter are quickly obtained from the trained neural networks. The reduced solution is given by the linear combination of the reduced basis functions with the
	modal coefficients as weights. The computational cost of the 
	online phase is much smaller than the cost of the offline phase. 
\end{itemize}

Next, we are going to describe the building blocks of the above algorithm.


\subsubsection{Proper orthogonal decomposition}

The POD procedure is widely used to compress data \cite{Bergmann2009,Baiges2014,Burkardt2006,hesthaven2016,benner2020}. We adopt it to extract an orthonormal basis with a least squares approach \cite{Kunisch2002,Quarteroni2015}.

Let us consider the discrete finite dimensional set $\{\bm \mu_1, \dots,\bm \mu_{{L}} \} \subset \mathcal{P}$. 
The FOM is solved for each parameter value $\bm \mu_l, l=1,\dots, L$, and the corresponding high fidelity solutions $\bar \Phi=\{ \bm{\bar v}, \bm{\bar p}, \bm{\bar u}, \bm{\bar w}, \bm{\bar q}\}$ 
called snapshots, are arranged in the snapshot matrix $S_{\Phi}$ 
as follow: \begin{equation*}
    {S}_{\Phi} = \begin{Bmatrix}
    \bar \Phi_1(\bm{\mu}_1) & \cdots & \bar \Phi_1(\bm{\mu}_{L}) \\ 
    \vdots & \vdots & \vdots  \\ 
    \bar \Phi_{{N}_{\delta}}(\bm{\mu}_1) & \cdots &  \bar \Phi_{{N}_{\delta}}(\bm{\mu}_{L})
    \end{Bmatrix}.
\end{equation*}
The POD space can be extracted by solving the following eigenvalue problem \cite{Kunisch2002}:
\begin{equation*}
    C_{\Phi} \bm c_l = \lambda_l \bm c_l, \quad l = 1, \dots, L,
\end{equation*}
where $C_{\Phi}=\frac{1}{L}S_{\Phi}^TS_{\Phi}\in \mathbb{R}^{L\times L}$ is the correlation matrix associated with the snapshots. For fixed $R \ll L$, the reduced orthonormal basis is constructed as follow:
\begin{equation*}
    \bm w_r = \frac{1}{\sqrt{\lambda_r}} S_{\Phi}\bm c_r, \quad r = 1,\dots, R.
\end{equation*} 
The value of $R$ is commonly chosen to reach a user-provided threshold $\varepsilon$ for the  cumulative energy of the eigenvalues:
\begin{equation}
\frac{\sum_{i=1}^{R}\lambda_i^2}{\sum_{i=1}^{L}\lambda_i^2} \ge \varepsilon.
\label{eq:energy}
\end{equation}
 
Let us store the vectors of the reduced basis functions (also called modes) in matrix $V_\delta$: 
\begin{equation}
    {V_\delta}=\left[ \bm{w}_1| \dots |\bm{w}_R \right] \in \mathbb{R}^{ N_{\delta} \times R}.
\end{equation}
Then, the reduced solution $\Phi_{\text{r}} (\bm \mu_l)$ can be written as:
\begin{equation}
 \Phi_{\text{r}} (\bm \mu_l) = \sum_{r=1}^{R} ({V_\delta}^T  \bar \Phi (\bm \mu_l))_r \bm w_{r}, \quad l=0,\dots,L,
\end{equation}
where $({V_\delta}^T  \bar \Phi(\bm \mu_l))_r$ is the modal coefficient associated with the $r$-th mode. 
Note that the POD space coincides with the resolution of the following minimization problem \cite{Kunisch2002,Quarteroni2015}:
\begin{equation}
    \min_{{V_\delta}} \Vert {S}_{\Phi}-{V_\delta}{V_\delta}^T{S}_{\Phi} \Vert \quad s.t. \quad {V_\delta}^T{V_\delta}=I,
    \label{min}
\end{equation}
which optimizes the distance between the snapshots and their projection onto the reduced space.

\subsubsection{Neural network interpolation} 

\begin{figure}[htb]
\centering
\begin{tikzpicture}[x=2.4cm,y=1.2cm]
  \readlist\Nnod{2,4,4,2} 
  \readlist\Nstr{n,m,k} 
  \readlist\Cstr{i,h^{(\prev)},o} 
  \def\yshift{0.55} 
  
  \foreachitem \N \in \Nnod{
    \def\lay{\Ncnt} 
    \pgfmathsetmacro\prev{int(\Ncnt-1)} 
    \foreach \i [evaluate={\c=int(\i==\N); \y=\N/2-\i-\c*\yshift;
                 \x=\lay; \n=\nstyle;
                 \index=(\i<\N?int(\i):"\Nstr[\n]");}] in {1,...,\N}{ 
      \node[node \n] (N\lay-\i) at (\x,\y) {$\strut\Cstr[\n]_{\index}$};
      
      \ifnumcomp{\lay}{>}{1}{ 
        \foreach \j in {1,...,\Nnod[\prev]}{ 
          \draw[white,line width=1.2,shorten >=1] (N\prev-\j) -- (N\lay-\i);
          \draw[connect] (N\prev-\j) -- (N\lay-\i);
        }
        \ifnum \lay=\Nnodlen
          \draw[connect] (N\lay-\i) --++ (0.5,0); 
        \fi
      }{
        \draw[connect] (0.5,\y) -- (N\lay-\i); 
      }
      
    }
    \path (N\lay-\N) --++ (0,1+\yshift) node[midway,scale=1.6] {$\vdots$}; 
  }
  
  \node[above=3,align=center,mydarkgreen] at (N1-1.90) {Input\\[-0.2em]layer};
  \node[above=2,align=center,mydarkblue] at (2.5,1.50) {Hidden\\[-0.2em]layers};
  \node[above=3,align=center,mydarkred] at (N\Nnodlen-1.90) {Output\\[-0.2em]layer};
  
\end{tikzpicture}

\flushleft\caption{Feedforward neural network.}
\label{my_net}
\end{figure}
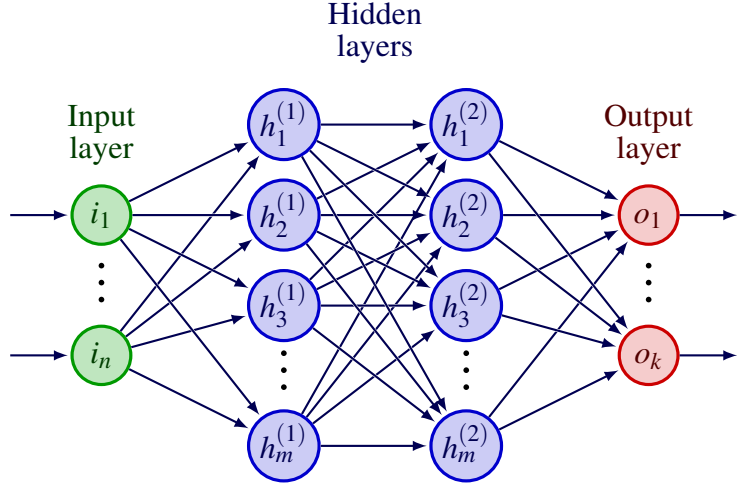

We choose a fully connected feedforward neural network \cite{Goodfellow2016,Kriesel2007,Calien2020}, also called perceptron. Such network is a set of layered neurons (or nodes), where each neuron of a layer is connected with all the neurons of the next layer through oriented edges (or synapses) \cite{Rosenblatt1958,Minsky1969,Fine2006}. The input layer takes the set of parameters 
and the output layer returns the modal coefficients. 
Others neurons form the hidden layers (see Figure \ref{my_net}). 

During the training process in the offline phase, we optimize the weights of the synapses with
the backpropagation algorithm \cite{Rumelhart1986,Rojas1996}. A loss function, which measures the distance between the actual output and the required output, is minimized by computing (backward) the gradient with respect to the weights. Hyperparameters such as the activation function, the number of layers, the number of neurons per layer and the learning rate
are tuned to improve and accelerate the learning process (see \cite{Sharma2017,montana1989} for further details). The values resulting from this tuning procedure are reported in Sec.~\ref{sec:num_res}.

The neural network offers an approximation ${\bm \pi}^{\rm NN}$ of the function:
\begin{equation}\label{eq:pi}
    \bm \pi: \bm \mu_l \mapsto [({V_{\delta}}^T  \bar \Phi (\bm \mu_l))_r]_{r=1}^R, \quad l=1,\dots,L.
\end{equation}
Once the network is trained, the reduced solution can be computed for every new parameter value $\bm \mu_{\text{new}}$ during the online phase as \cite{Chen2021,Hesthaven2018,Pichi2021,Shah2021}: 
\begin{equation}
\Phi_{\text{r}} (\bm \mu_{\text{new}}) = \sum_{r=1}^{R} \pi^{\mathrm{NN}}_r(\bm \mu_{\text{new}}) \bm w_{r},
\end{equation}
where $\pi^{\mathrm{NN}}_r$ is the $r$-th component of $\bm \pi^{\mathrm{NN}}$. 

\section{Numerical results}\label{sec:num_res}




In order to compare our methodology with the one proposed in \cite{zakia2021}, we consider the same numerical setting, which is briefly recalled in the following. Using TetGen \cite{hangtet}, we build a triangular mesh with a given size 
over the boundary $\partial\Omega$ and a tetrahedral mesh  of the same size inside the volume $\Omega$. 
The total number of elements for the generated mesh is 42354. 
The inf-sup stable 
$\mathbb{P} 2-\mathbb{P} 1$ elements are employed
for velocity and pressure. 

Let $\bm t_c$ be the vector tangent to the vessel centerline (obtained using VMTK as explained in \cite{zakia2021}) in axial direction, $R$ the maximum vessel radius (with respect to the centerline) and $r$ the distance between the mesh nodes and the centerline.
We defined the desired blood flow velocity in the entire domain $\Omega$ 
as
\begin{equation*}
      \bm v_{\mathrm{m}} = \vconst\Big(1-\frac{r^2}{R^2}\Big)\bm t_c,
\end{equation*}
where $\vconst=350\,\mms$ is the desired velocity magnitude. 
At the inlets, we prescribe the following velocity:
\begin{equation}\label{eq:vinlet}
    \bm v_{in} = \frac{\eta Re}{R_{in}}\Big(1-\frac{r^2}{R_{in}^2}\Big)\bm n_{in},
\end{equation}
where $\eta=3.6\,\mmqs$ is the constant kinematic viscosity, $Re$ is the Reynolds number, 
$R_{in}$ is the maximum radius of the given inlet surface, and $\bm n_{in}$ denotes the outward unit normal to the inlet surface.
Notice the inlet velocity profiles are parameterized by 
the Reynolds number in \eqref{eq:vinlet}, which is the variable parameter for the study presented in this Section, while the ``measured'' velocity $\bm v_{\mathrm{m}}$ is independent of $Re$. In fact, we want to address the case when $\bm v_{\mathrm{m}}$ comes from in-vivo flow measurements and the associated Reynolds number is expected to vary in a certain range. Our optimal control problem will identify the normal stress that has to be imposed at the outlet in order for the computed velocity field to be the best approximation of $\bm v_{\mathrm{m}}$ for each chosen $Re$ in the expected range.

Recalling that  $\bar\Phi=\{ \bar{\bm v}, \bar{ \bm p}, \bar{\bm u}, \bar{\bm w}, \bar{\bm q}\}$ is the \emph{truth solution} and $\Phi_{\rm r}=\{ \bm v_{\rm r}, \bm p_{\rm r}, \bm u_{\rm r}, \bm w_{\rm r}, \bm q_{\rm r} \}$ is the \emph{reduced solution}, we introduce the absolute and relative $L^2$ errors defined as
\begin{equation}\label{eq:errore}
    \epsilon^{\mathrm{abs}}_{\Phi} = \norm{\bar\Phi-\Phi_{\rm r}}_{L^2}, \qquad
    \epsilon^{\mathrm{rel}}_{\Phi} = \frac{\epsilon^{\mathrm{abs}}_{\Phi}}{\norm{\bar\Phi}_{L^2}},
\end{equation}
respectively.
The total relative error is defined as:
\begin{equation}\label{eq:erroreTot}
\begin{split}
    \mathscr{E}^{\mathrm{rel}} &= \displaystyle\frac{\big((\epsilon^{\mathrm{abs}}_{\bm v})^2+(\epsilon^{\mathrm{abs}}_{p})^2+(\epsilon^{\mathrm{abs}}_{\bm u})^2+(\epsilon^{\mathrm{abs}}_{\bm w})^2+(\epsilon^{\mathrm{abs}}_{q})^2\big)^{1/2}}{(\norm{\bar{\bm v}}_{L^2}^2 +\norm{\bar{\bm p}}_{L^2}^2+ \norm{\bar{\bm u}}_{L^2}^2+ \norm{\bar{\bm w}}_{L^2}^2+ \norm{\bar{\bm q}}_{L^2}^2)^{1/2}}.
\end{split}
\end{equation}

Following \cite{zakia2021},
we let the Reynolds number vary in the interval $[70,80]$. The choice of this interval is to ensure that the flow remains laminar everywhere in the domain,
to demonstrate the feasibility of our approach, and to compare its performance with the method in \cite{zakia2021}. 
We consider equispaced values 
$\mu_l=Re$, for $l=1,\dots,L$ with $L=100$ as in \cite{zakia2021}.
We split the finite dimensional set of parameters into a train set $\mutrain$ and a test set $\mutest$. The former is used during the offline phase to build the reduced basis space and to train the feedforward neural networks, while 
the latter is used during the online phase to validate the method by reconstructing the solutions and comparing them with the corresponding snapshots. For our tests, we choose $\mutrain=\{\mu_{2m-1}\}$ and $\mutest=\{\mu_{2m}\}$, with $m=1,\dots,L/2$. 

\begin{figure}[h]
    \centering
    \includegraphics[width=0.48\columnwidth]{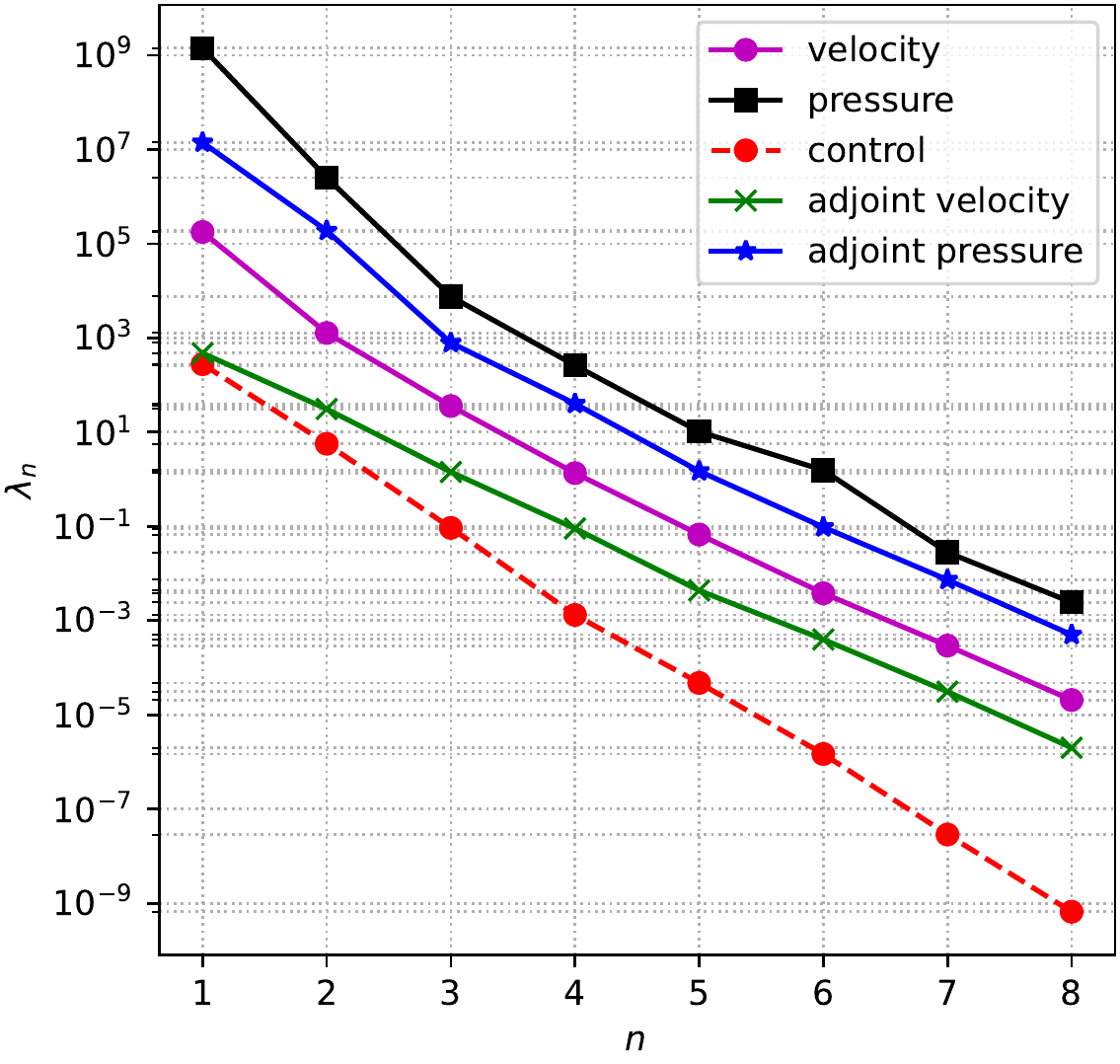}
    \caption{Eigenvalues of the snapshots matrix for all the variables in our problem.}
    \label{fig:eigenvalues}
\end{figure}

The solutions associated with all the parameters $\mu_l$, with $l=1,\dots,L$, are computed during the offline phase using Python libraries FEniCS \cite{logg2012,aln2015} and \textit{multiphenics}  \cite{ballarinmultiphenics2019}. This represents the snapshots set. Then, we consider subset $\bm \mu_{train}$ to build the snapshots matrix $S_{\Phi}$ and the corresponding reduced space through the POD algorithm as explained in Sec.~\ref{sec:rom}.
In Figure \ref{fig:eigenvalues}, we show the eigenvalues associated with the POD on the snapshots matrix $S_{\Phi}$ for all variables using $n=8$ reduced basis functions. 
We see that for every $n$ the largest eigenvalue is associated with the pressure, while the smallest eigenvalue is associated with the control variable. In addition, the eigenvalues for the velocity and pressure snapshots matrices are larger than the eigenvalues of the respective adjoint counterpart. 
Table \ref{tab:cumulative_energy} reports the cumulative energy \eqref{eq:energy} of the first three eigenvalues. We observe that more than the 99.99\% of the energy of the eigenvalues is retained with the first two eigenvalue for all the variables.

\begin{table}[h]
    \centering
    \begin{tabular}{c|r|r|r}\toprule
        Variable & \multicolumn{1}{c|}{$\lambda_1$} & \multicolumn{1}{c|}{$\lambda_2$} & \multicolumn{1}{c}{$\lambda_3$}\\\midrule
        Velocity & 99.99482\% &  100\% & \multicolumn{1}{c}{-} \\
        Pressure & 99.99968\% &  100\% & \multicolumn{1}{c}{-} \\
        Control & 99.95826\% &  99.99999\% &  100\% \\
        Adjoint velocity & 99.57621\% &  99.99910\% &  100\%\\
        Adjoint pressure & 99.98195\% &  100\% & \multicolumn{1}{c}{-} \\\bottomrule
    \end{tabular}
    \caption{Cumulative energy associated with the first three eigenvalues.}
    \label{tab:cumulative_energy}
\end{table}

Once the POD step is completed, the offline phase 
is concluded by training the neural networks to compute the modal coefficients. 
The entire ROM has been implemented within Python package EZyRB \cite{ezyrb2018}, developed and maintained by the mathLab group at SISSA.
For the neural networks, we choose the Sigmoid activation function $f(x) = (1+e^{-x})^{-1}$ and we consider two 
fully connected hidden layers with the same number of neurons.
Since we have obtained satisfactory results with two hidden layers, we have not increased the number of layers. 
We have estimated the 
optimal number of neurons per layer and the optimal learning rate 
of the neural network 
through an exhaustive search. We varied the number of neurons in $\{20,30,\dots,100\}$ and the learning rate in $\{1\text{e-}01,1\text{e-}02,\dots,1\text{e-}05\}$ for each variable and for a different number of reduced basis functions. The results of this search are shown in Table \ref{tab:nn}. As we can see, the 
optimal number of neurons per layer and the optimal learning rate vary widely from one variable to the other and for different numbers of reduced basis functions. There seems to be no way to predict the optimal network configuration, which makes the exhaustive search by trial and error unavoidable. The results reported next have been obtained with the optimal neural networks.


\begin{table}[h]
    \centering
    \begin{tabular}{c|c|c|c|c|c|c|c|c|c}\toprule
        \multirow{2}{*}{Variable} & \multirow{2}{*}{Parameter} & \multicolumn{8}{c}{Number of basis functions}\\
        & & 1 & 2 & 3 & 4 & 5 & 6 & 7 & 8 \\\midrule
        \multirow{2}{*}{Velocity} & Neurons/layer & 90 & 50 & 100 & 40 & 60 & 70 & 20 & 90\\
        & Learning rate & 1e-05 & 1e-05 & 1e-04 & 1e-05 & 1e-04 & 1e-03 & 1e-05 & 1e-03\\\hline
        \multirow{2}{*}{Pressure} & Neurons/layer & 20 & 50 & 100 & 30 & 30 & 60 & 80 & 20 \\
        & Learning rate & 1e-05 & 1e-05 & 1e-05 & 1e-05 & 1e-04 & 1e-05 & 1e-03 & 1e-04\\\hline
        \multirow{2}{*}{Control} & Neurons/layer & 40 & 30 & 40 & 70 & 70 & 70 & 20 & 20 \\
        & Learning rate & 1e-05 & 1e-05 & 1e-05 & 1e-03 & 1e-04 & 1e-05 & 1e-02 & 1e-05\\\hline
        Adjoint & Neurons/layer & 50 & 90 & 50 & 30 & 70 & 30 & 40 & 30 \\
        velocity & Learning rate & 1e-05 & 1e-05 & 1e-04 & 1e-04 & 1e-05 & 1e-05 & 1e-03 & 1e-03\\\hline
        Adjoint & Neurons/layer & 20 & 40 & 70 & 60 & 60 & 40 & 90 & 70 \\
        pressure & Learning rate & 1e-05 & 1e-05 & 1e-05 & 1e-05 & 1e-02 & 1e-02 & 1e-05 & 1e-05\\
        \bottomrule
    \end{tabular}
    \caption{Optimal values of neurons per layer and learning rate for each variable and for different numbers of reduced basis functions.}
    \label{tab:nn}
\end{table}

\begin{figure}[h]
    \centering
    \includegraphics[width=0.48\columnwidth]{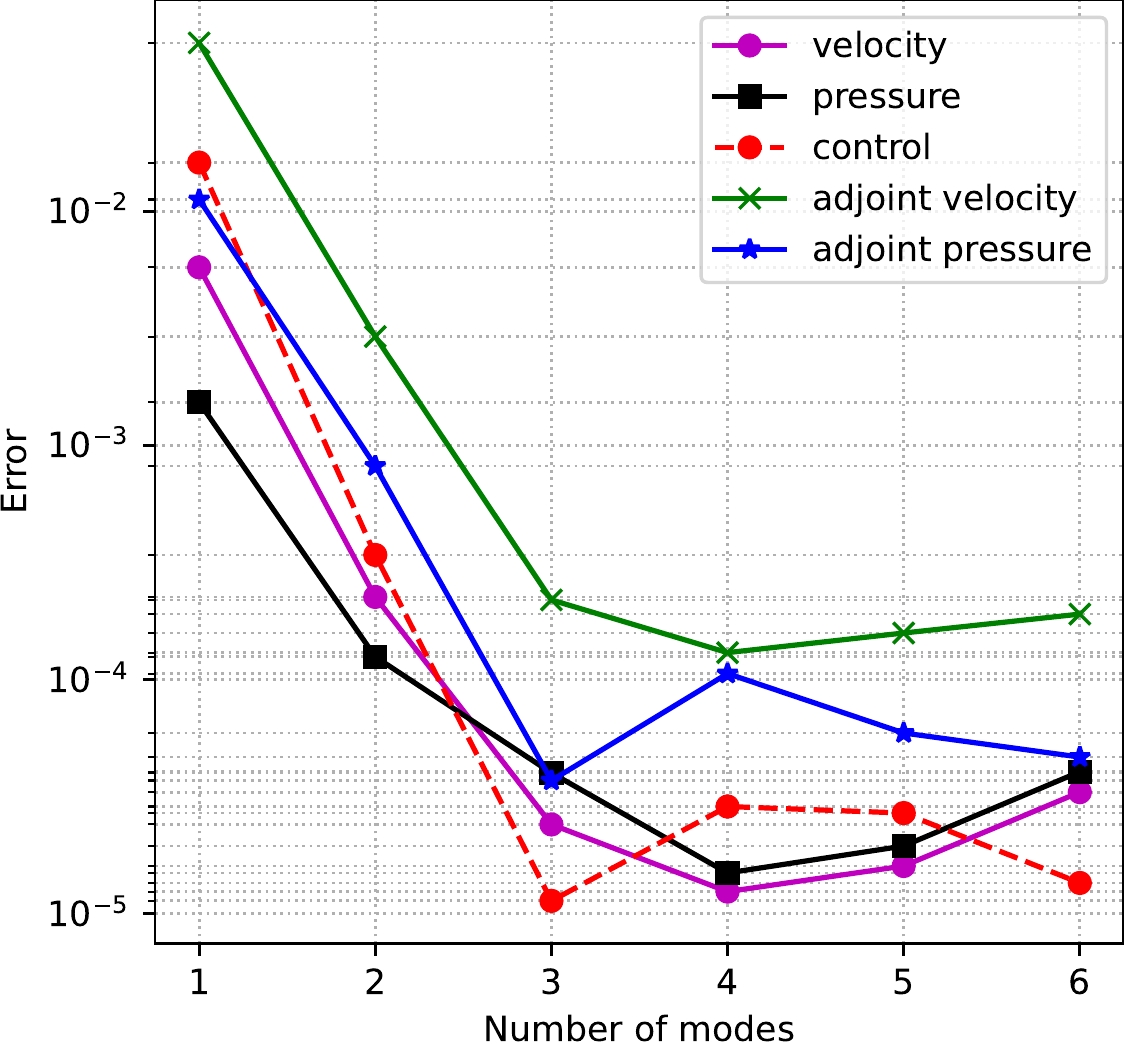}\quad
    \includegraphics[width=0.47\columnwidth]{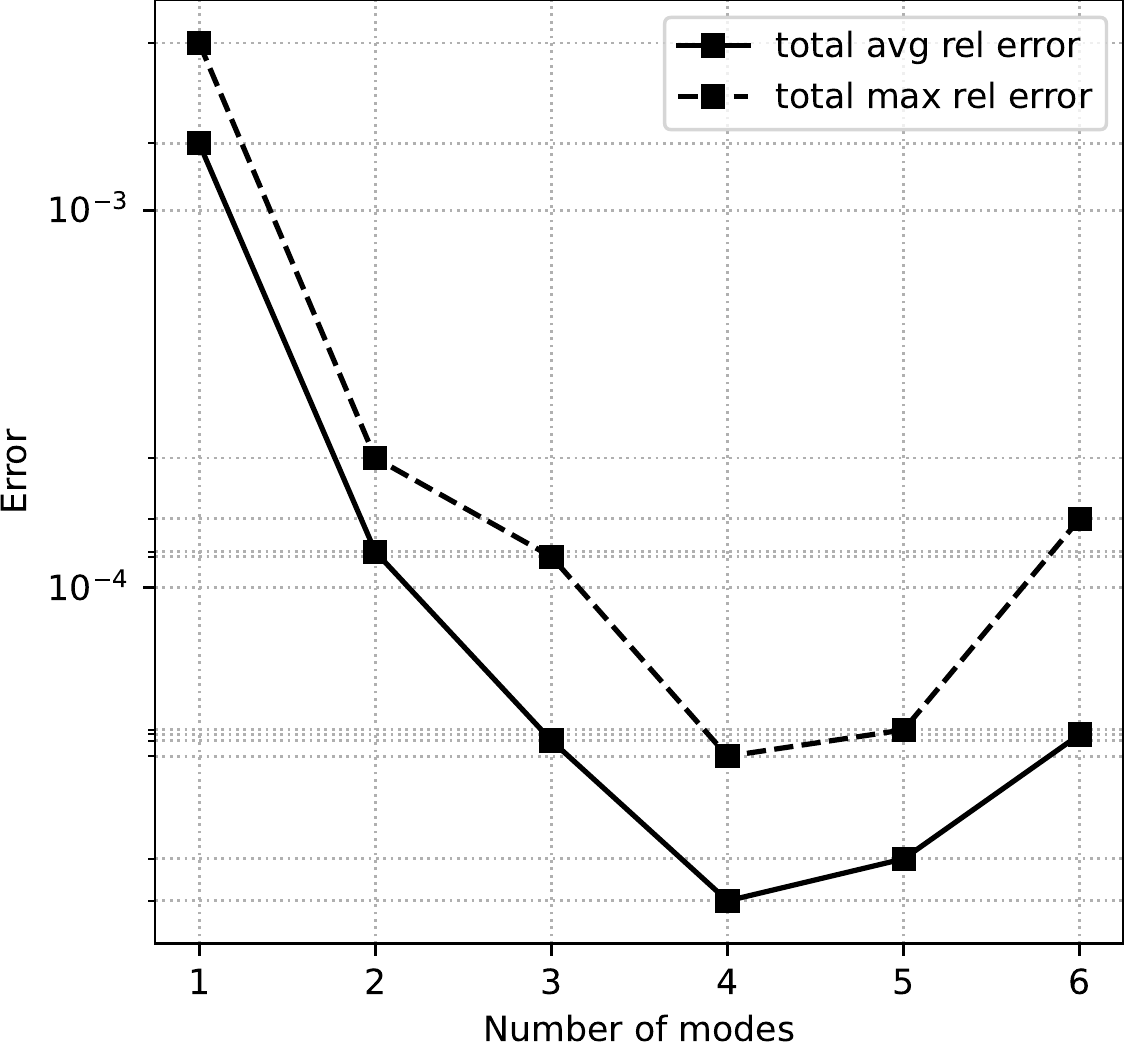}
    \caption{Left: average relative error \eqref{eq:errore} for each variable as the number of reduced basis functions is increased. Right: total maximum and average relative error \eqref{eq:erroreTot} as the number of reduced basis functions is increased.}
    \label{fig:error_comparison}
\end{figure}

In the online phase, we use the trained neural networks to construct the approximated solutions corresponding to the parameters in $\mutest$. Figure \ref{fig:error_comparison} (left) shows the average relative error \eqref{eq:errore} for each variable as the number of reduced basis functions $n$ is increased and Figure \ref{fig:error_comparison} (right) displays the maximum and average relative error \eqref{eq:erroreTot} as $n$ varies. 
We note that the errors reach their minimum for $n=3$ or $n=4$, depending on the variable. 
For almost every $n$, larger average relative errors are attained for the adjoint variables.
The average order of accuracy is about $\mathscr{O}(10^{-4})$ when using more than one reduced basis function, which compares well with \cite{zakia2021}.

We conclude this section by fixing $n=6$ in order to compare our results with those in \cite{zakia2021}. We start by reporting the computational times required by our methodology
in Table \ref{tab:time}. During the offline phase, about 540\,s are need to compute each snapshot with the full order model, while the POD and the training of a neural network require about 18\,s. 
The average computational time
required by the online phase 
for all the cases in $\mutest$
is $\mathscr{O}(10^{-4})$ 
leading to an average speed-up of $\mathscr{O}(10^{6})$, which is 4 times larger than the speed-up obtained using the POD-Galerkin approach in \cite{zakia2021}.
This is a remarkable improvement.

\begin{table}[h]
    \centering
    \begin{tabular}{c|c|c|c}\toprule
        \multicolumn{3}{c|}{Offline phase} & \multirow{2}{*}{Online phase} \\
        FOM & POD & NN train & \\\midrule
        540 & 0.06 & 17 & 9.2e-04
        \\\bottomrule
    \end{tabular}
    \caption{Computational times in s required by the offline phase (broken into sub-phases) and the online phase of our ROM. 
    }
    \label{tab:time}
\end{table}

\begin{figure}
    \centering
    \begin{subfigure}[b]{0.4\textwidth}
        \centering
        \includegraphics[scale=0.23]{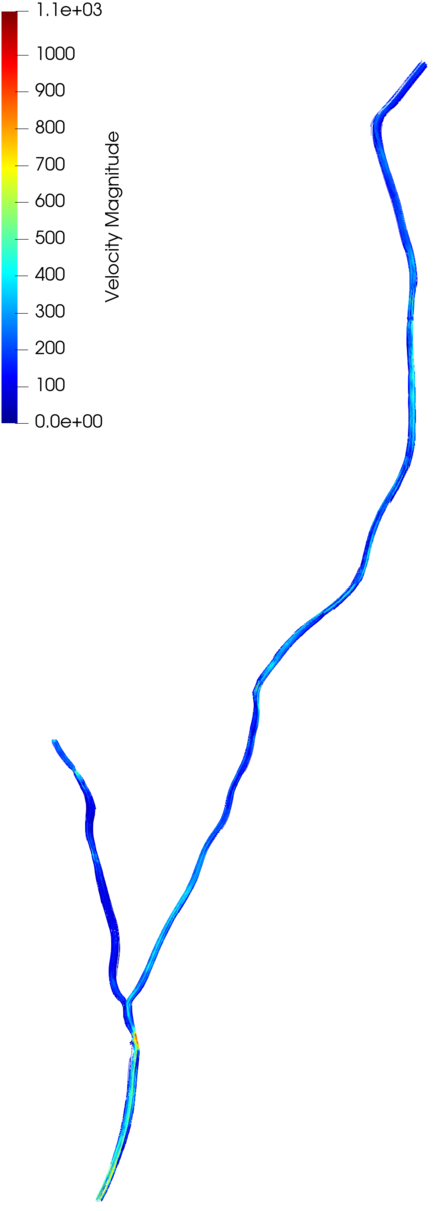}
        \caption{ROM velocity}
        \label{fig:plots_comparison:a}
    \end{subfigure}
    \quad
    \begin{subfigure}[b]{0.3\textwidth}
        \centering
        \includegraphics[scale=0.23]{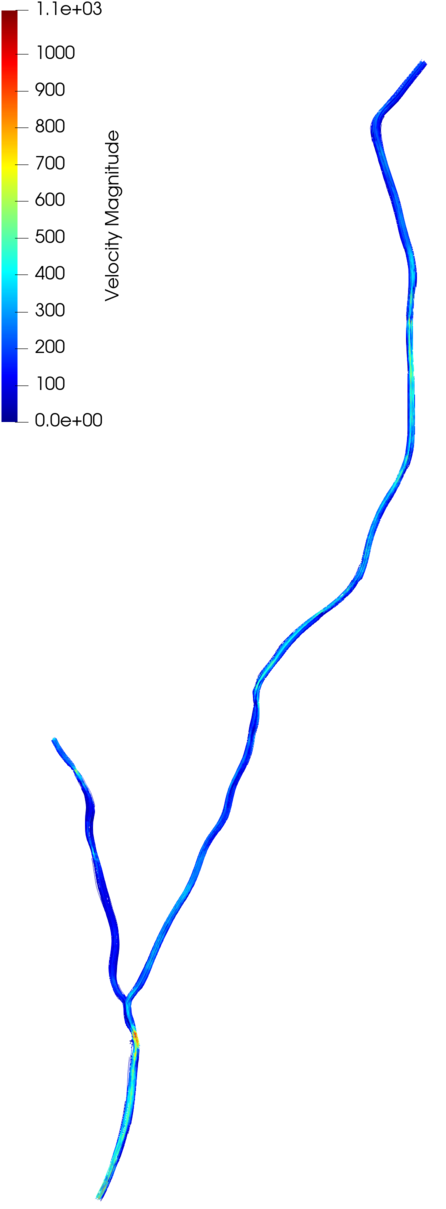}
        \caption{FOM velocity}
        \label{fig:plots_comparison:b}
    \end{subfigure}
    \quad
    \begin{subfigure}[b]{0.4\textwidth}
        \centering
        \includegraphics[scale=0.23]{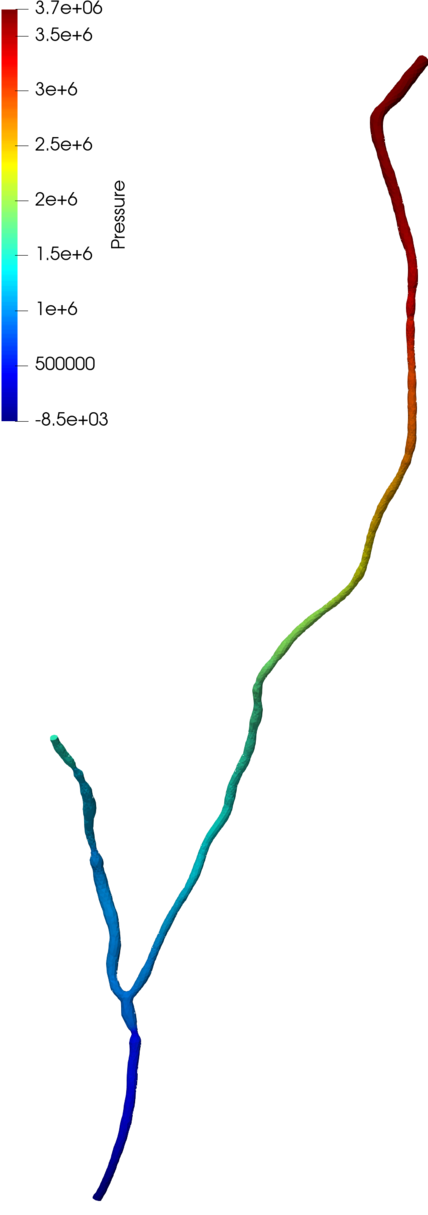}
        \caption{ROM pressure}
    \end{subfigure}
    \begin{subfigure}[b]{0.4\textwidth}
        \centering
        \includegraphics[scale=0.23]{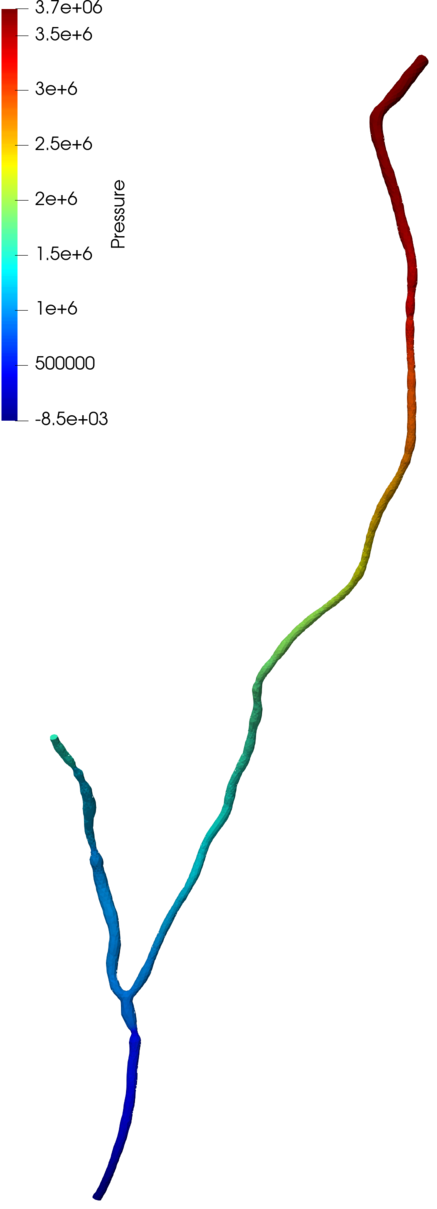}
        \caption{FOM pressure}
    \end{subfigure}
    \caption{Comparison of the velocity (mm/s) field computed by (A) ROM and (B) FOM
    and pressure (mm$^2$/s$^2$) field given by the (C) ROM
   and (D) FOM for $Re=80$. The number of POD modes used for the ROM is $ n = 6$.}
    \label{fig:plots_comparison}
\end{figure}

\begin{figure}
    \centering
    \begin{subfigure}[b]{0.38\textwidth}
        \centering
        \includegraphics[scale=0.15]{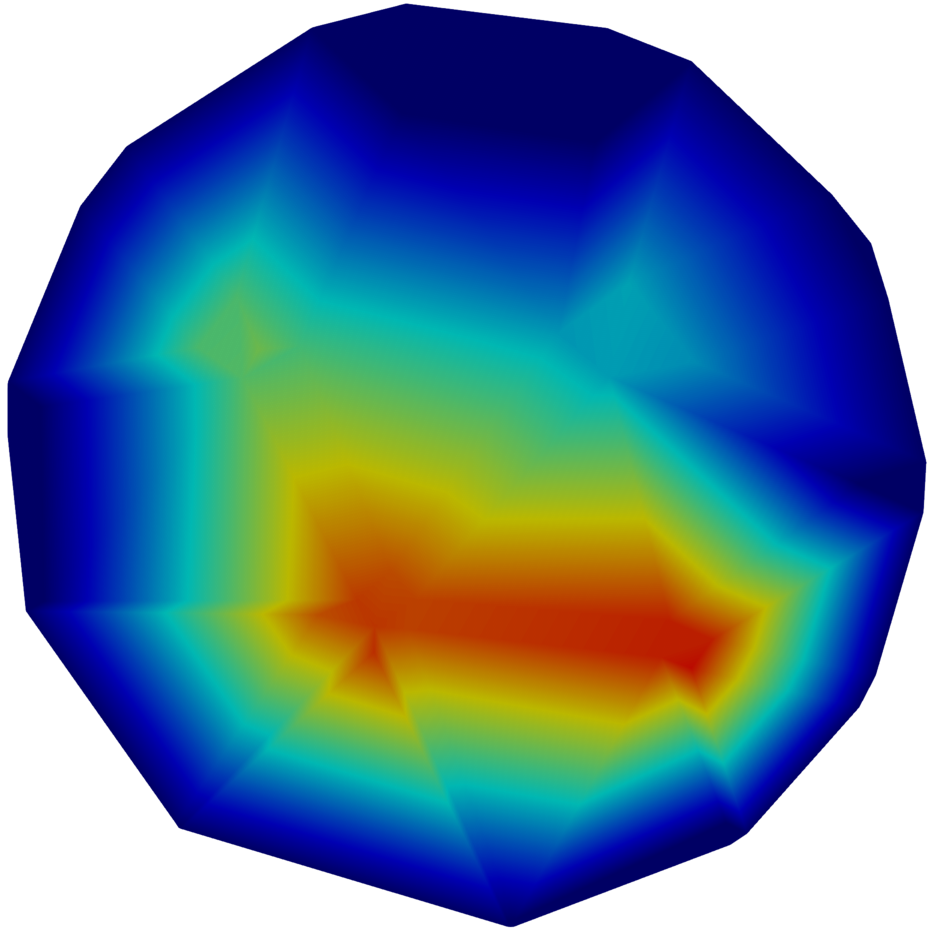}
        \caption{ROM control}
    \end{subfigure}
    \qquad
    \begin{subfigure}[b]{0.38\textwidth}
        \centering
        \includegraphics[scale=0.165]{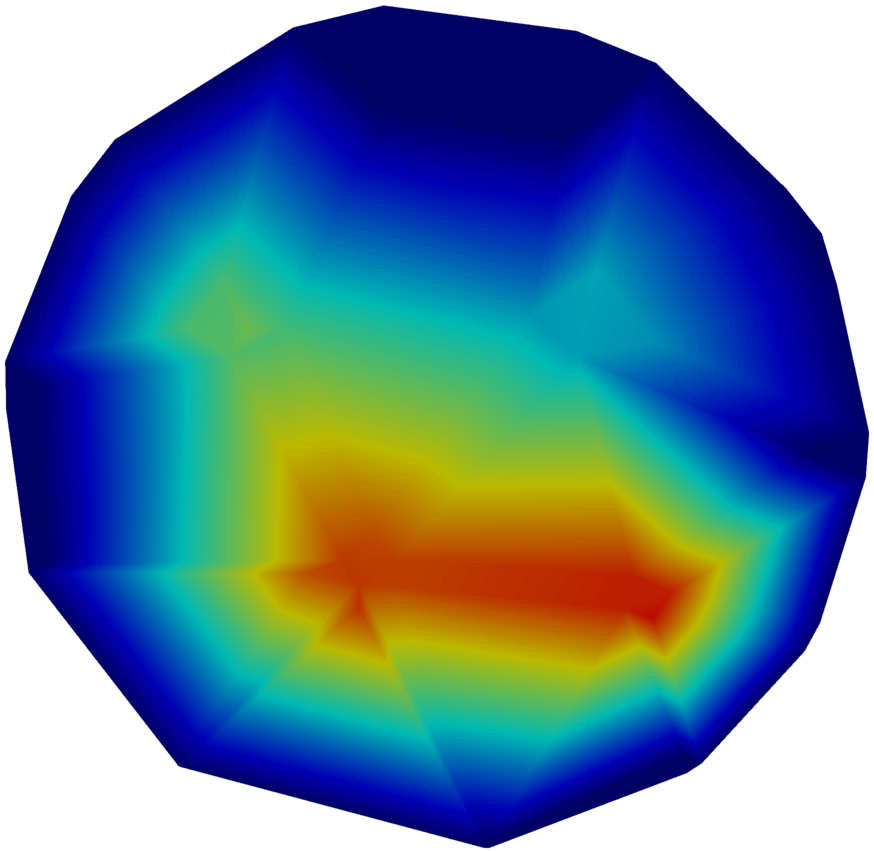}
        \caption{FOM control}
    \end{subfigure}
    \includegraphics[scale=0.13, trim = 0cm -8cm 0cm 0cm]{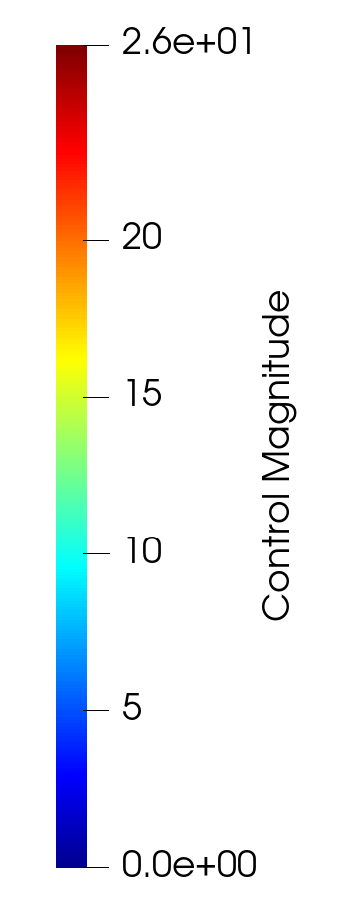}
    \caption{
    Control variable (mm$^2$/s$^2$) at the outflow boundary for $Re=80$ computed by the (A) ROM and (B) FOM. The number of POD modes used for the ROM is $ n = 6$.}
    \label{fig:control_comparison}
\end{figure}

Figure \ref{fig:plots_comparison} shows a qualitative comparison between the velocity and pressure fields for $Re = 80$ computed by ROM and FOM. We see a substantial qualitative agreement over the entire geometry for both flow variables. In particular, notice that our ROM approach is able to provide a good approximation for the velocity also in the region downstream the graft, where the magnitude is higher due to the flow rate introduced by the RITA  (compare Figure \ref{fig:plots_comparison} (A) with \ref{fig:plots_comparison} (B)).
Figure \ref{fig:control_comparison} reports the ROM-FOM comparison for the control variable, again for $Re = 80$.
We observe very good agreement also between ROM-computed and FOM-computed control variable
over the entire outlet surface. 

\section{Conclusions and future perspectives}\label{sec:concl}
We focused on a parameterized optimal blood flow control problem in a patient-specific coronary artery bypass graft geometry. The bypass was performed with the right internal thoracic artery on the left anterior descending artery. The optimal control variable is the normal stress at the outlet needed to set the outlet boundary condition in order to minimize in the least squares sense the difference between approximated and measured velocity.
The blood flow is modeled by the steady-state incompressible Navier-Stokes equations and parameterized by the Reynolds number. We restricted our attention to a physiological range of Reynolds numbers that generate laminar flow. 

To reduce the high computational costs of a finite element method (FEM) for our parameterized optimization problem, we introduced a non-intrusive reduced order method based on the so-called POD-NN approach. During the offline phase of our ROM, three steps are performed: (i)
the high fidelity solutions are computed via FEM and collected into the snapshots matrix, (ii) proper orthogonal decomposition is performed on the snapshots matrix to build a basis for the reduced order space, and (iii) a neural network is trained to estimate the modal coefficients for each variable. After this expensive phase, we were able to simulate the hemodynamics corresponding to the desired Reynolds number in a considerably reduced time
during the online phase. 
Our numerical tests show that this data-driven methodology is as accurate as the POD-Galerkin approach proposed in \cite{zakia2021}, but it is considerably faster. 
In particular, we managed to achieve a speed-up of about 4 orders of magnitude, i.e., the simulation is almost real-time.

The proposed framework is a preliminary step in the direction of building a user-friendly virtual platform that could be used for clinical studies \cite{girfogliononintrusive}. The quantification of the normal stress at the outlet can be improved with the introduction of surrogate lumped parameter network models. This would introduce additional parameters that could be found through an automated quantification via the control variables as done in \cite{girfoglpodirom}. This work is limited to one physical parameter (i.e., the Reynolds number). The introduction of geometrical features, such as the degree of the stenosis, would expand the set of possible configurations and generalize our approach to a wider range of clinical cases as shown in \cite{sienadata}. Finally, following \cite{toll2013,strazz2021} one could increase the realism and complexity of the problem by making it time-dependent and accounting for the compliance of the arterial wall.

\vskip 6mm
\noindent{\bf Acknowledgements}

\noindent
We acknowledge the support provided by the European Research Council Executive Agency by the Consolidator Grant project AROMA-CFD ``Advanced Reduced Order Methods with Applications in Computational Fluid Dynamics" - GA 681447, H2020-ERC CoG 2015 AROMA-CFD, PI G. Rozza, and INdAM-GNCS 2019-2020 projects.
This work was also partially supported by US National Science Foundation through grant DMS-1953535 (PI A.~Quaini). A.~Quaini acknowledges 
support from the Radcliffe Institute for Advanced Study at Harvard University where she has been the 2021-2022 William and Flora Hewlett Foundation Fellow.

\bibliographystyle{abbrv}
\bibliography{biblio.bib}

\end{document}